\documentclass[
 reprint,
 amsmath, amssymb,
 pre,
 aps,
 superscriptaddress
]{revtex4-2}

\usepackage{graphicx}
\usepackage{subfigure}
\usepackage{hyperref}
\usepackage{bm}
\usepackage{float}
\usepackage{color}
\usepackage{amsmath}
\usepackage{bbm}
\usepackage{braket} 
\usepackage{xcolor}

\definecolor{darkred}{RGB}{205,92,92}

\newcommand{\abs}[1]{\left\vert {#1} \right\vert}

\newcommand{\expval}[1]{\braket{#1}}
\hypersetup{colorlinks=true, citecolor=blue, urlcolor=blue, linkcolor=blue}
\begin{document}

\preprint{APS/123-QED}

\title{Quantum predator-prey cycles in dissipative Rydberg array}

\author{Ya-Xin Xiang}
    \affiliation{National Laboratory of Solid State Microstructures and School of Physics,
	Collaborative Innovation Center of Advanced Microstructures, Nanjing University, Nanjing 210093, China}

\author{Weibin Li}
    \affiliation{School of Physics and Astronomy, and Centre for the Mathematics and Theoretical Physics of Quantum Non-equilibrium Systems, University of Nottingham, Nottingham, G7 2RD, UK}

\author{Zhengyang Bai}
     \email{zhybai@nju.edu.cn}
     \affiliation{National Laboratory of Solid State Microstructures and School of Physics,	Collaborative Innovation Center of Advanced Microstructures, Nanjing University, Nanjing 210093, China}
	
\author{Yu-Qiang Ma}
     \email{myqiang@nju.edu.cn}
     \affiliation{National Laboratory of Solid State Microstructures and School of Physics,	Collaborative Innovation Center of Advanced Microstructures, Nanjing University, Nanjing 210093, China}	 \affiliation{Hefei National Laboratory, Hefei 230088, China}

\begin{abstract}
The predator-prey cycle is a paradigmatic example of self-organized population oscillations in far-from-equilibrium systems, yet testing its underlying mechanism in natural ecosystems is often precluded by uncontrollable microscopic parameters.
Here, we propose a quantum analogue of predator-prey dynamics using a tunable two-dimensional Rydberg atom array.
Through mean-field analysis and large-scale numerical simulations based on the open-system discrete truncated Wigner approximation, we demonstrate stable predator-prey cycles of Rydberg excitations on microsecond timescales. 
We show that nonperturbative quantum coherence drives spontaneous time-translation symmetry breaking, leading to globally synchronized oscillations under short‑range interactions. 
Even under desynchronization caused by quantum jumps, locally synchronized predator-prey cycles persist, with amplitude scaling inversely with the square root of the system size.
Our work extends the study of predator-prey models to the quantum realm and advances quantum simulation strategies that leverage engineered many-body nonequilibrium effects.
\end{abstract}

\maketitle
\section{Introduction}
The predator–prey cycle is a paradigmatic example of spontaneous population oscillations in far‑from‑equilibrium systems, classically described by deterministic models such as the Lotka–Volterra equations~\cite{volterra1926fluctuations,lotka1920analytic}.
Moving beyond mean-field (MF) descriptions, spatial stochastic lattice models incorporate discrete individuals, local interactions, and demographic noise, revealing rich phenomena—such as pattern formation and noise-induced transitions~\cite{mobilia2007phase,mckane2005predator,tauber2011stochastic}—that lie outside the scope of continuum models. 
More broadly, the core concept that simple local interactions can produce emergent oscillations has inspired studies of spontaneous oscillations across diverse fields, including chemical reactions~\cite{bray1921periodic,prigogine1968symmetry,Zhabotinsky1973auto}, biological rhythms~\cite{tyson2003sniffers,tyson2001network,goldbeter2002comp}, epidemic spread~\cite{bartlett1957measles,greer2020emergence,Hethcote1989}, and evolutionary game theory~\cite{sigmund1989oscillations,sinervo1996the,roman2013interplay}.

Despite this theoretical progress, empirical validation of predator–prey cycles in natural ecosystems remains challenging due to the difficulty of isolating and controlling key parameters—such as reproduction rates, interaction ranges, and intrinsic noise—in real-world settings.
This limitation motivates the search for clean, tunable platforms in which the essential ingredients of far‑from‑equilibrium population dynamics can be implemented and probed with microscopic control. 

Atomic systems excited to high-lying Rydberg states have emerged as a powerful platform for investigating nonequilibrium quantum many-body dynamics~\cite{sibalic2018rydberg,weimer2010rydberg,browaeys2020many,saffman2010,lin2025AI, chenSpectroscopy}.
The strong and programmable Rydberg-Rydberg interactions drive a rich array of self-organization phenomena~\cite{ebadi2021quantum,semeghini2021probing,bernien2017probing,liang2025,li2024Observation, manovitz2025quantum}, enabling a bottom-up understanding of how macroscopic collective behavior originates from microscopic quantum interactions. 
Dissipative dynamics—engineered through spontaneous decay and controlled dephasing—further enables the realization of far-from-equilibrium phases inaccessible in closed systems~\cite{wu2024dissipative,helmrich2020signatures,ding2020phase,ding2024ergodicity, Bleuenn2025controlled, russo2025Quantumb}.

Recently, classical Lotka–Volterra-type dynamics have been observed in thermal Rydberg vapors~\cite{jiao2024Quantuma}, where strong dephasing and ionization induce a dissipative long‑range coupling between ions and Rydberg atoms.
These experiments showcase the possibility of realizing predator-prey cycles in atomic systems. Thermal vapors operate in the classical regime, where quantum coherence is washed out, leaving open the question of how genuine quantum effects—fluctuations and nonperturbative coherence—shape such dynamics. Cold Rydberg atoms in optical tweezer arrays offer a complementary platform with single-atom control, scalability, programmable interactions, and tunable dissipation, enabling precise exploration of both coherent and dissipative mechanisms at the quantum level~\cite{weimer2010rydberg,browaeys2020many,saffman2010,lin2025AI, chenSpectroscopy}.  A fully quantum theoretical investigation of predator-prey dynamics—especially in the coherent regime of Rydberg arrays, and the role of interaction range and quantum fluctuations in stabilizing such oscillations—remains an open frontier. 

\begin{figure}[t]
    \centering
	\includegraphics[width=1.\columnwidth, keepaspectratio]{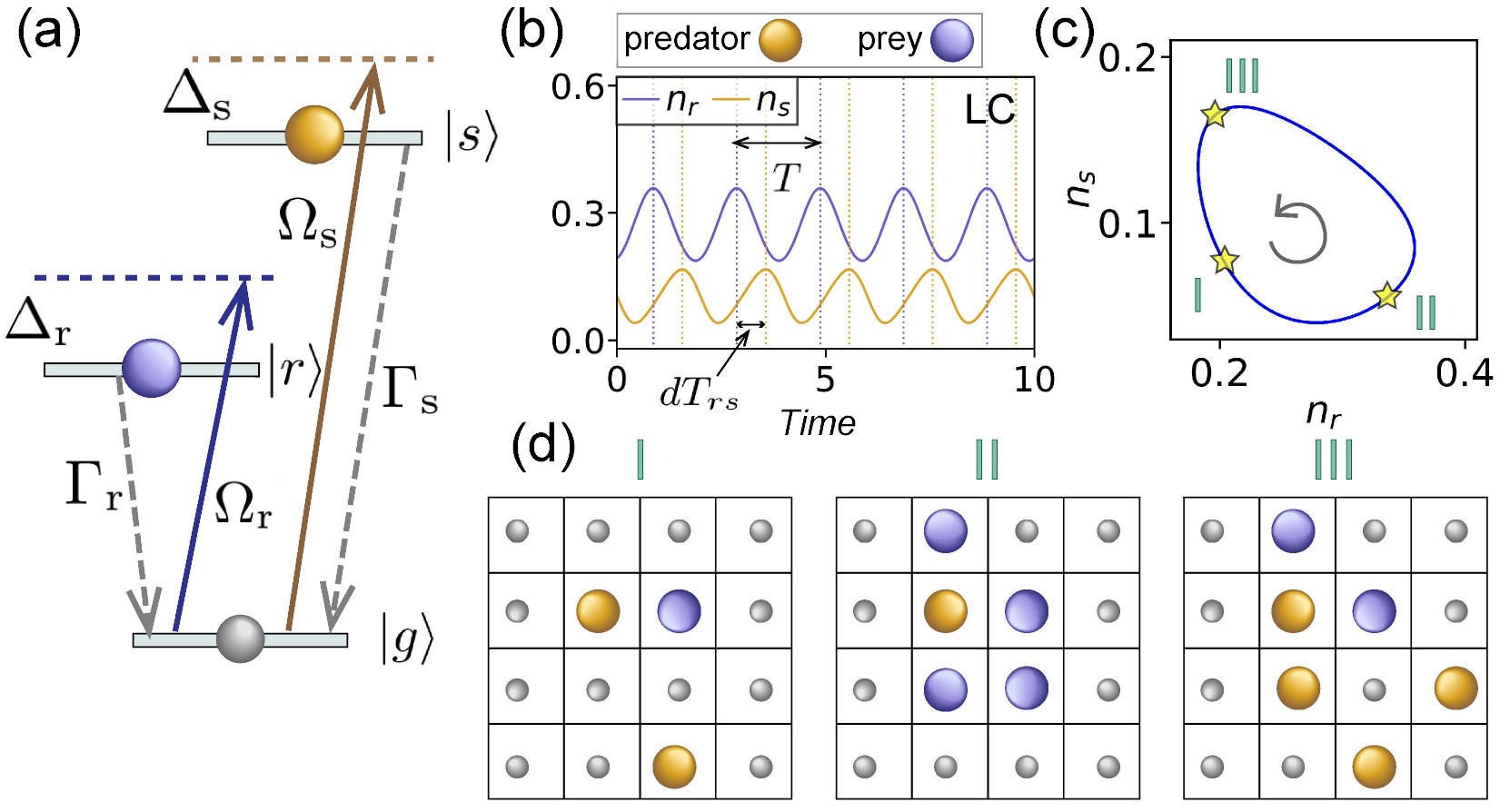}
	\caption{(a) Atomic-level scheme. The ground state $\ket{g}$ is coupled to the Rydberg states $\ket{s}$ and $\ket{r}$ by a laser field with Rabi frequencies $\Omega_\text{s}$, $\Omega_\text{r}$ and detunings $\Delta_s$, $\Delta_r$.
    A nonzero relative detuning $\Delta_r-\Delta_s$ introduces asymmetry in the effective interactions between the two Rydberg components.
    Excited atoms in the state $\ket{s}$ ($\ket{r}$) decay to the ground state at rates $\Gamma_r$ ($\Gamma_s$). 
    Mean-field  analysis reveals (b) predator-prey dynamics with alternating dominance between $\ket{r}$ (preys) and $\ket{s}$ (predators) populations characterized with period $T \approx 2/\Gamma_r$ and relative time advance $dT_{rs}$ of peaks in $n_r$ (yellow dotted lines) over peaks in $n_s$ (purple dotted lines), manifesting a 
    (c) limit cycle (LC)  in the $n_s$-$n_r$ phase space with
    (d) three distinct dynamical phases emerging subsequently:
    In phase I, low excited populations favor $\ket{g}\to\ket{r}$ transition, while interaction-induced level shifts suppress $\ket{g}\to\ket{s}$ transition. 
    In phase II, high $n_r$ enhances $\ket{g}\to\ket{s}$ transition and suppress $\ket{g}\to\ket{r}$ transition.
    In phase III, both excitation processes are blocked and the system returns to the low-excitation phase I. }
	\label{fig:model}
\end{figure}

Here, we demonstrate the emergence of predator-prey cycles in a two-component two-dimensional (2D) Rydberg
atom array, as schematized in Figs.~\ref{fig:model}(a) and (b). The predator-prey limit cycle (LC)  manifests as periodic out-of-phase oscillations between one component ($\ket{s}$, predator) and the other ($\ket{r}$, prey) and evolves through three distinct dynamical states: (I) prey accumulation, (II) subsequent predator growth, and (III) suppression of both [see Figs.~\ref{fig:model}(c) and (d)]. Crucially, these LCs emerge exclusively in the quantum regime without classical counterparts, as the dynamics of quantum coherence is indispensable for breaking the time-translation symmetry. 

Using the open-system discrete truncated Wigner approximation (OSDTWA), we reveal quantum-jump-induced quasicycles—slow oscillatory modes whose oscillation amplitude scales inversely with the square root of the system size. Notably, these quantum quasicycles exhibit frequencies distinct from their classical counterparts~\cite{mckane2005predator,mobilia2007phase}, suggesting a fundamentally different microscopic mechanism for noise amplification.
In small systems, the predator-prey LCs remain robustly synchronized  under finite-range van der Waals (vdW) interactions. As the system size increases, however, local quantum jumps induce significant desynchronization,
suppressing global oscillation, while local synchronization persists.
This persistence suggests that a large system fragments into numerous spatio-temporal clusters, each sustaining prey-predator cycles locally—a signature resolvable in current Rydberg atom array experiments~\cite{ manovitz2025quantum}. The hidden predator-prey relation in large systems is further revealed through two-time cross-correlation functions, offering a clear window into the underlying nonequilibrium dynamics.

\section{Model}

We study a spin-1 model for a dissipative Rydberg gas, as illustrated in Fig.~\ref{fig:model}(a).
It comprises $N$ atoms arranged in a 2D array where the ground state $\ket{g}$ is coupled to two highly-excited Rydberg levels $\ket{r}$, $\ket{s}$ by lasers with Rabi frequencies $\Omega_r$, $\Omega_s$ and detuning $\Delta_r$, $\Delta_s$ from resonance.
Upon excitation, the Rydberg states $\ket{r}$ and $\ket{s}$ interact via strong and long-range interactions. 
This system permits a microscopic description via a quantum master equation for the many-body density operator $\hat{\rho}$ ($\hbar=1$ hereinafter),
\begin{equation}
\partial_t \hat{\rho} = i\left[\hat{\rho},\hat{H}\right] + \hat{\boldsymbol{D}}[\hat{\rho}],
\end{equation}
where the Hamiltonian $\hat{H}$ responsible for the coherent dynamics reads
\begin{equation}
\begin{aligned}
\hat{H}=&\sum_{l=1}^N{\left[ \left( \frac{\Omega_s}{2} \hat{\sigma}_l^{gs} + \frac{\Omega_r}{2}\hat{\sigma}_l^{gr} +\text{h.c.}\right) - \Delta_r \hat{\sigma}_l^{rr} -  \Delta_s \hat{\sigma}_l^{ss} \right]}\\
&+\sum_{l,k\neq l}{ \left(V_{lk,ss}\hat{\sigma}_k^{ss} + V_{lk,sr}\hat{\sigma}_k^{rr}\right)\hat{\sigma}_l^{ss} }\\
&+\sum_{l,k\neq l}{\left(V_{lk,rr} \hat{\sigma}_k^{rr} + V_{lk,sr}\hat{\sigma}_k^{ss}\right)\hat{\sigma}_l^{rr}},\notag
\end{aligned}
\end{equation}
where h.c. stands for Hermitian conjugation, and $l, k$ denotes atomic sites and the transition operator $\hat{\sigma}^{ab}_l\equiv\ket{a}_l\bra{b}_l$ for $a,b=g,s,r$, and the interactions between Rydberg states $\ket{a}$ and $\ket{b}$ are parameterized by $V_{lk,ab}$. 
The dissipative dynamics arising from  decay of the Rydberg states $\ket{r}$ and $\ket{s}$ to the ground state is described by
$\hat{\boldsymbol{D}}[\hat{\rho}]=\sum_{l, a=s,r}{\Gamma_a \left(\hat{\sigma}_l^{ga}\hat{\rho}\hat{\sigma}_l^{a g} - \left\{\hat{\sigma}_l^{aa},\hat{\rho}\right\}\right/2)}$ with the decay rate $\Gamma_a$~\cite{Bleuenn2025controlled, russo2025Quantumb}. 
The Rydberg-Rydberg interaction terms $V_{kl,ab}\hat\sigma_k^{aa}\hat\sigma_l^{bb}$ introduce a density-dependent effective detuning. This density dependence serves as the origin of the nonlinearity, which gives rise to the LCs observed in our systems.

\section{Results}

\subsection{Mean-field analysis}
To characterize the dynamical behavior of the system, we begin with a MF  ansatz by assuming that the total density matrix factorizes as $\hat{\rho} = \prod_l \hat{\rho}_l$, with interactions treated via a MF coupling ($a,b=s,r$) $\chi_{ab}=2N^{-1}\sum_{k\neq l} V_{lk, ab}$. 
The phase diagram is determined by the fixed points in the equations of motion for the expectation values ${\sigma}^{ab} = N^{-1}\sum_l \langle \hat{\sigma}^{ab}_l \rangle$ and the eigenvalues of their corresponding $8\times8$ Jacobian.
The two-component Rydberg populations are denoted by $n_a = {\sigma}^{aa}$. Throughout this study, we adopt a parameter set with $\Gamma_s = \Gamma_r = \Gamma = 1$, $\Omega_s = \Omega_r = \Omega$, $\Delta_s = 8$, and $\chi_{ab} = \chi = 12$ as a representative case.

\begin{figure}[t]
	\centering
	\includegraphics[width=1.\columnwidth, keepaspectratio]{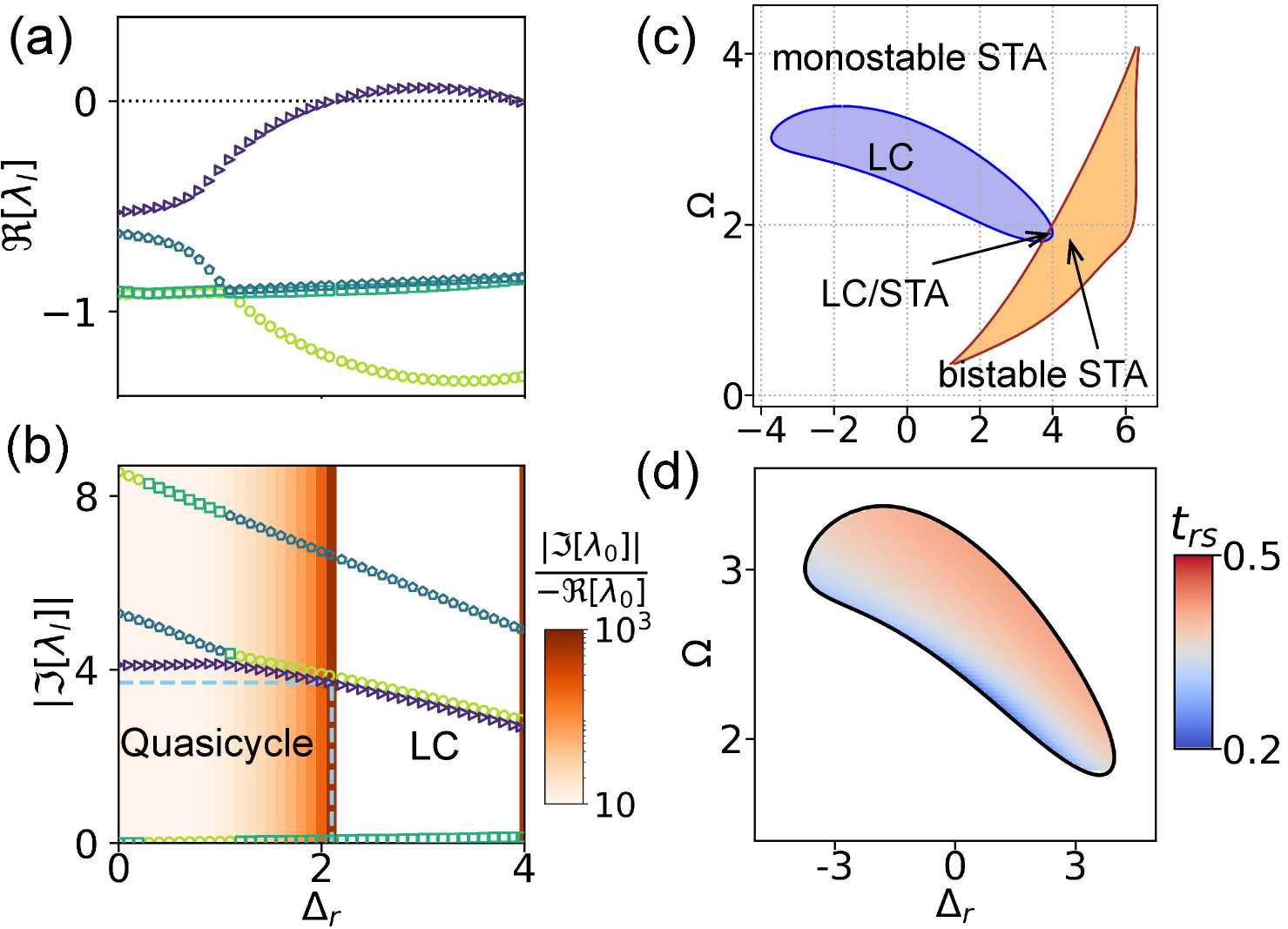}
	\caption{Mean-field analysis. (a) Real and (b) imaginary parts of the Jacobian eigenvalues $\lambda_l$. Quasicycles occur when the ratio (color-coded) $-\abs{\Im[\lambda_0]}/\Re[\lambda_0]\gg 1$. In panel (b), the blue dashed lines mark a quasicycle at $\Delta_r=2.1$ with frequency $\approx 3.7$. Eigenvalues are conjugate pairs sorted by real parts in descending order.  (c) Phase diagram in $\Omega$-$\Delta_r$ parameter space showing monostable stationary (STA), purely LC, STA/LC coexistence, and bistable STA phases. (d) Normalized time advance $t_{rs}=dT_{rs}/T$ of LC, where  $dT_{rs}$ is the time advance of $n_r$ peaks relative to $n_s$ and $T$ is the period. Parameters for (a)(b):  $\Omega=2$.}
	\label{fig:MF}
\end{figure}

The eigenvalues, displayed as complex conjugate pairs and sorted by their real parts $\Re[\lambda_0]\geq\Re[\lambda_1]\geq...$ are shown in Figs.~\ref{fig:MF}(a) and (b). The real and imaginary parts of $\lambda_0$ govern the relaxation timescale and oscillation frequency of fluctuations near the fixed point, respectively.  A fixed point is linearly stable if $\Re[\lambda_0] < 0$ and unstable if $\Re[\lambda_0] > 0$.
For small values of the detuning $\Delta_r$, a stable fixed point occurs. Notably, the real parts of the dominant eigenvalue pair $\lambda_0, \lambda_0^*$ increase with $\Delta_r$, transitioning from negative to positive values [Fig.~\ref{fig:MF}(a)], signifying a dynamic phase transition from a stationary state to a LC state. 

Near the transition, quasicycles emerge when the condition $-\abs{\Im[\lambda_0]}/\Re[\lambda_0]\gg 1$ is satisfied [Fig.~\ref{fig:MF}(b)]. 
The distinction between a LC and a quasicycle is determined by the linear stability of the fixed point. 
A LC arises from linear instability, corresponding to $\Re[\lambda_0] > 0$. 
In contrast, a quasicycle corresponds to a long-lived oscillatory relaxation mode, characterized by $\Re[\lambda_0] < 0$.
For quasicycles, the oscillation amplitude decays exponentially at rate $-\Re[\lambda_0]$; thus, to exhibit sustained oscillatory behavior over many cycles, the imaginary part $\abs{\Im[\lambda_0]}$, which sets the oscillation frequency, must be much larger then the decay rate.

The phase diagram spanned by $\Omega$ and $\Delta_r$ features four distinct regimes: monostable stationary state, purely LC state, coexistence between two stationary states and a LC state, and bistable stationary states [Fig.~\ref{fig:MF}(c)].
In the LC region, the oscillation occurs with a period on the order of microseconds, set by the atomic decay rate~\cite{sibalic2018rydberg,wu2024dissipative}.
The emergent predator-prey cycles are characterized by the normalized time advance $t_{rs}=dT_{rs}/T$ [see Fig.~\ref{fig:model}(b)], representing the relative timing between peaks in $n_s$ and $n_r$ within an oscillation period $T$.

Owing to the periodicity of the oscillations, $t_{rs}$ is constrained to the interval $[-0.5, 0.5)$. A value of $t_{rs}=0$ corresponds to synchronization between the two Rydberg components, while positive values in the range $t_{rs} \in (0.2, 0.5)$ [see Fig.~\ref{fig:MF}(d)] indicate a predator-prey dynamic. In this regime, atoms in state $\ket{s}$ act as predators that ``feed'' on those in state $\ket{r}$. 
This emergent nonreciprocity originates from the asymmetric effective interactions between the two Rydberg states, intrinsically driven by the relative detuning $\Delta_r - \Delta_s$ (where $\Delta_r \simeq 2\sim4$ and $\Delta_s = 8$ in our calculations). 
Such a parameter disparity results in state-dependent effective interactions, fundamentally distinguishing this quantum mechanism from the stochastic contact processes typically observed in nonreciprocal classical systems~\cite{mobilia2007phase,mckane2005predator,tauber2011stochastic}. Notably, the predator-prey roles are exactly reversed upon the exchange of detunings $\Delta_r \leftrightarrow \Delta_s$.

We point out that quantum coherence plays an important role in the emergence of LC and quasicycles. To demonstrate this, we analyze the rate equation derived from the MF equations~\cite{lesanovsky2013Kinetic, helmrich2020signatures} [see Sec.~I of the Supplementary Material (SM)]. The resulting Jacobian has purely real negative eigenvalues, indicating that the rate equation supports only stable fixed points. In other words, both the LC and quasicycles in our system originate from quantum coherence and reside in the full eight-dimensional phase space.

\subsection{Two-dimensional Rydberg array}

\begin{figure}[b]
	\centering
	\includegraphics[width = 1.\columnwidth, keepaspectratio]{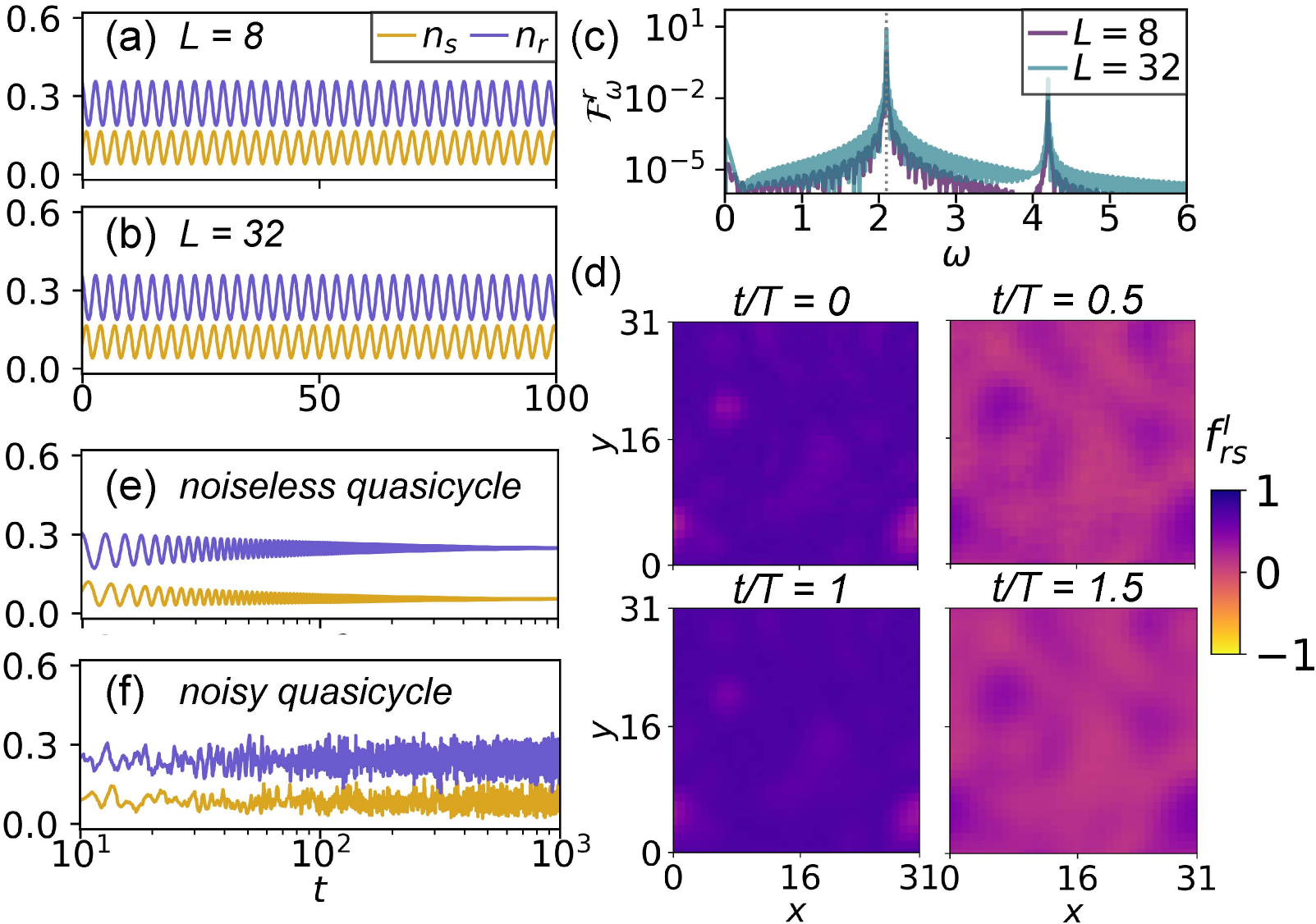}
	\caption{Results from OSDTWA with vdW coupling on a 2D array of edge length $L$. 
    Time series of average Rydberg population $n_s,n_r$ for both components in the LC phase for (a) $L=8$ and (b) $L=32$. 
    (c) The Fourier spectra for LC. Multiples of oscillation frequency are marked by gray dotted lines.
    (d) Spatial profile of the relative fraction $f_{rs}^l=(n_r^l-n_s^l)/(n_r^l+n_s^l)$ at half-period interval $T/2$. 
    The extrema $f_{rs}= 1$ and $f_{rs}=-1$ correspond to complete occupation of the $\ket{r}$ and $\ket{s}$ states, respectively 
    Time series of $n_s,n_r$ in the quasicycle state for $L=8$ (e) without and (f) with quantum noise.
    Parameters for (a)-(d): $\Delta_r=3$; parameters for (e)(f): $\Delta_r=2.1$. Only panel (f) includes quantum noise.}
	\label{fig:vdW_wo_jump}
\end{figure}

Our system is a 2D array of atoms comprising $N=L^{2}$ sites with unit spacing and edge length $L$, where each site is occupied by a single atom, consistent with experimental realizations~\cite{ebadi2021quantum,Guardado-Sanchez2018,manetsch2025tweezer}. 
For such an open quantum system,  the Liouville-space dimension grows exponentially as $\mathcal{O}(9^N)$, making exact numerical simulations challenging.
We employ the OSDTWA to simulate the stochastic spin-1 Rydberg array system. 
This approach combines the semiclassical framework of the discrete truncated Wigner approximation with the method of quantum jumps~\cite{singh2022driven,zhu2019generalized}
[see Sec.II of SM for details]. 
It is particularly well-suited for our system because it captures the quantum noise effects while remaining computational tractable for large system sizes, thus going beyond MF theory.

We first examine the deterministic evolution in the absence of quantum noise. In the MF LC regime, the spatially averaged Rydberg populations $n_a=\expval{n_a^l}_l$ ($a=r,s$) for both components display clear out-of-phase oscillations in both small [Fig.~\ref{fig:vdW_wo_jump}(a)] and large [Fig.~\ref{fig:vdW_wo_jump}(b)] systems. 
Their Fourier spectra $\mathcal{F}_\omega^a=\int{d\omega G_{aa}(t)e^{i\omega t}}$, where the two-time correlation functions~\cite{wu2024dissipative} ($a,b=r,s$)
\begin{equation}
\label{eq:G_t}
G_{ab}(t)=\expval{\delta n_a(t')\delta n_b(t+t')}_{t'}
\end{equation}
and $\delta n_a = n_a-\expval{n_a}_t$, shown in Fig.~\ref{fig:vdW_wo_jump}(c), exhibit sharp peaks at integer multiples of the LC frequency.
To further quantify the predator-prey dynamics, we introduce the relative fraction $f_{rs}^l =(n_r^l -n_s^l)/(n_r^l+n_s^l)$, $f_{rs}=\expval{f_{rs}^l}_l$, where $f_{rs} >0$ corresponds to dominance of the $r$-component atoms and $f_{rs}<0$ indicates dominance of the $s$-component atoms. The persistent prey-predator oscillations are reflected in the alternating spatial pattern of $f_{rs}^l$ across the array [Fig.~\ref{fig:vdW_wo_jump}(d)].

\begin{figure}[t]
	\centering
	\includegraphics[width=1.\columnwidth, keepaspectratio]{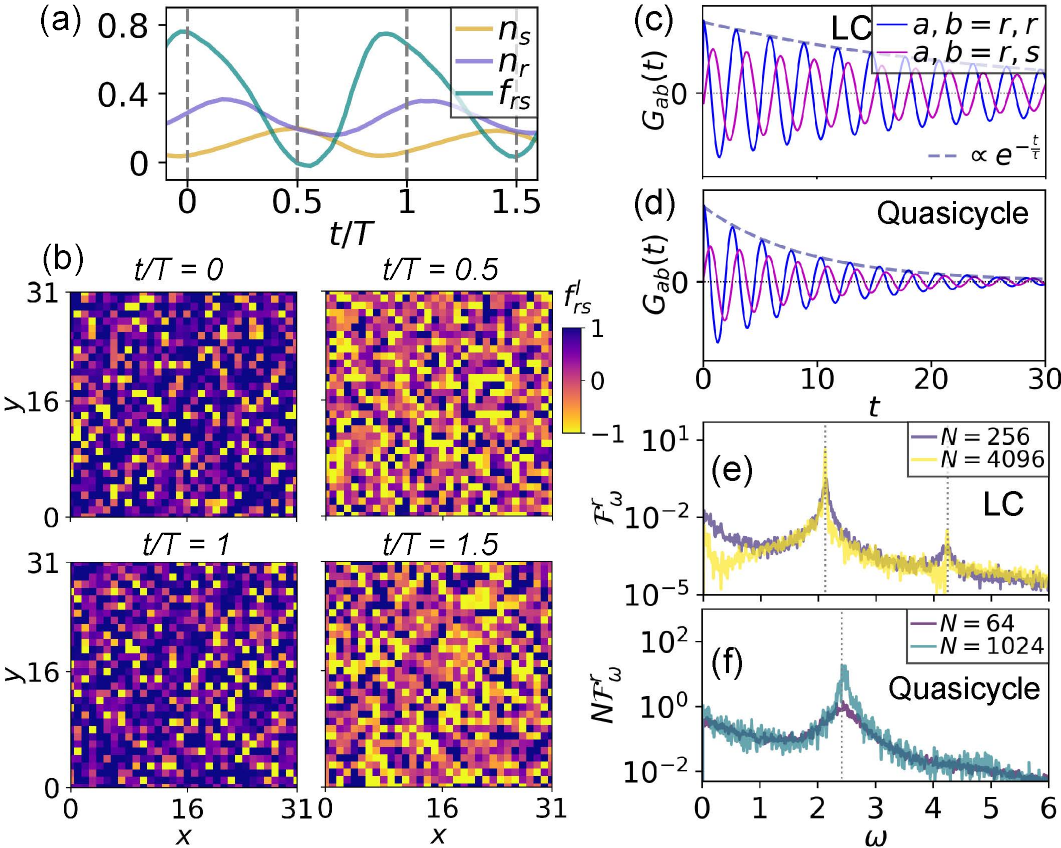}
	\caption{Predator-prey cycle under all-to-all coupling. (a) Time series of average Rydberg population $n_s,n_r$ for both components, and the relative fraction $f_{rs}=(n_r-n_s)/(n_r+n_s)$ for LC. Time is rescaled by the period $T$ measured from the time series, and the vertical dashed gray lines stand for half-period intervals, with the corresponding spatial profile of  $f_{rs}^l$ shown in (b). 
    The auto-correlation function $G_{rr}(t)=\expval{\delta n_r(t')\delta n_r(t'+t)}_{t'}$, and the cross-correlation function $G_{rs}(t)=\expval{\delta n_r(t')\delta n_s(t'+t)}_{t'}$ with $\delta n_{r,s}=n_{r,s}-\expval{n_{r,s}}_t$ for $N=256$ for (c) LC and (d) quasicycle. The characteristic lifetime $\tau$ is obtained from fitting the peak-envelope to $A e^{-t/\tau}$ (gray dashed lines).
    The Fourier spectra for (e) LC and  (f) quasicycles (rescaled by system size $N$). The intrinsic frequency is marked by gray dotted lines. Parameters for (a)-(c)(e): $\Delta_r=3$; parameters for (d)(f): $\Delta_r=2.1$. Correlation functions are averaged over trajectories of duration $\geq 10^4$ for well-converged statistics.}
	\label{fig:all_to_all}
\end{figure}

The separation of relaxation and oscillation timescales—
governed by the real and imaginary parts of the eigenvalues [Figs.~\ref{fig:MF}(a) and (b)]—indicates that fluctuations, whether from initial conditions or quantum noise, can induce quasicycles when the MF state is linearly stable.
This occurs when $\abs{\Im[\lambda_0]} \gg -\Re[\lambda_0]>0$, a condition satisfied by tuning the detuning parameter $\Delta$ from 3 to 2.1, which increases $\abs{\Im[\lambda_0]}$ and drives the transition from a LC to a quasicycle state.
Consistent with the MF prediction, the quasicycles decay slowly but are continuously regenerated by quantum noise [Figs.~\ref{fig:vdW_wo_jump}(e) and (f)].

While the system exhibits robust synchronization under vdW interactions, this coherence is vulnerable to quantum noise. All-to-all coupling—where each Rydberg atom interacts with a collective field generated by all other Rydberg atoms—counteracts this effect by suppressing atom-to-atom fluctuations. 
Within the parameter regime for MF LC, oscillatory dynamics persists [Fig.~\ref{fig:all_to_all}(a)]. However, local quantum jumps produce significant atom-to-atom fluctuations, as evidenced by the alternating spatial profile of $f_{rs}^l$ within the array [Fig.~\ref{fig:all_to_all}(b)].

The prey-predator dynamics is further confirmed with the oscillating auto-correlation function $G_{rr}$ and cross-correlation function $G_{rs}$ displayed in Fig.~\ref{fig:all_to_all}(c). The coherence time $\tau$ characterizes the decay of the envelope. The first peak of $G_{rs}$ locates at finite time delay $t$ indicates a phase shift between $r$- and $s$-components, consistent with the MF results shown in Fig.~\ref{fig:MF}(d). 
These oscillatory auto-correlation functions give rise to peaks in the Fourier components $\mathcal{F}_\omega^{r}$ at integer multiples of the intrinsic frequency, as depicted in Fig.~\ref{fig:all_to_all}(e).

\begin{figure}[t]
	\centering
	\includegraphics[width=1.\columnwidth, keepaspectratio]{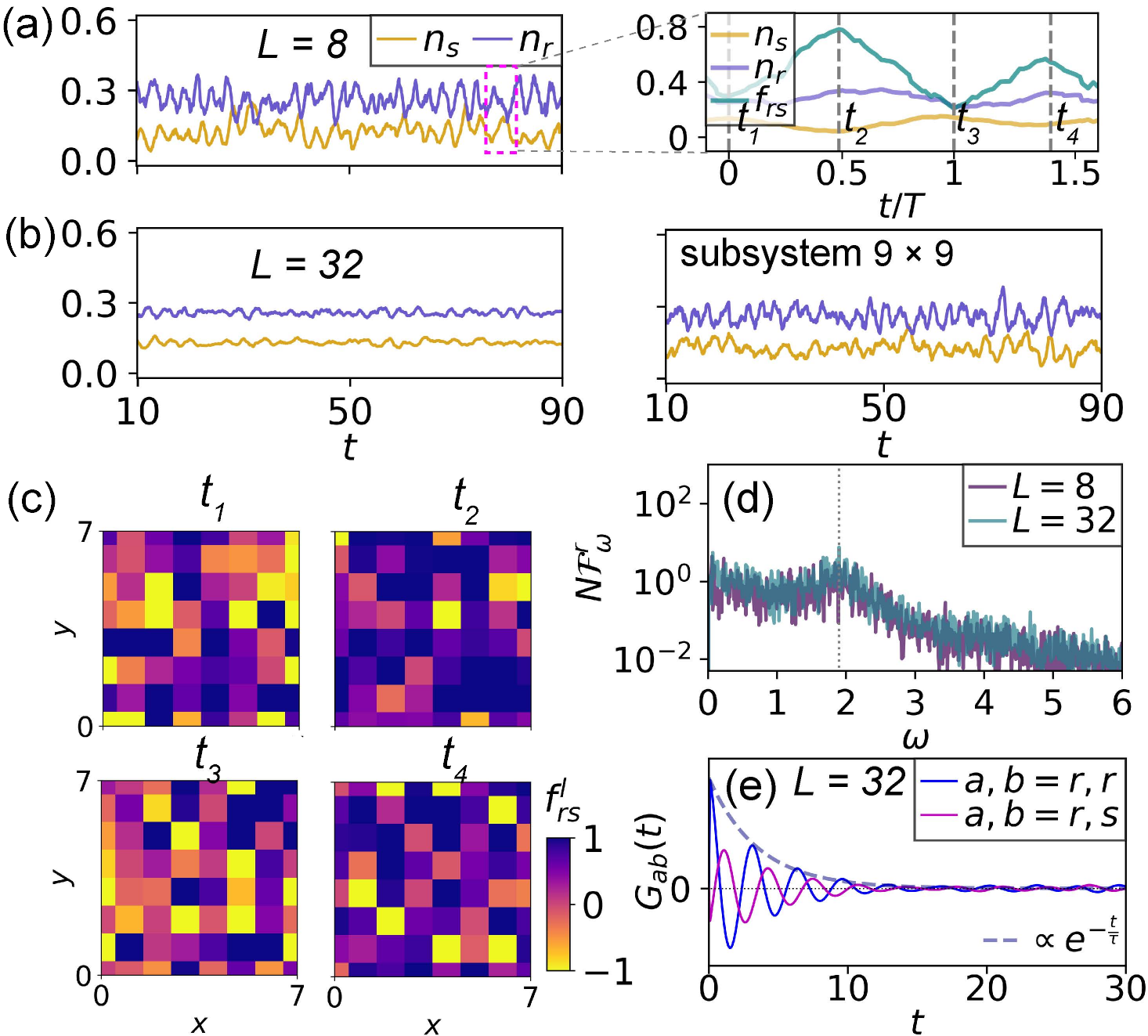}
	\caption{Locally synchronized predator-prey cycle with vdW coupling on a 2D lattice of edge length $L$. 
    (a) Time series of average Rydberg population $n_s$ and $n_r$ for a small system ($L=8$). Left: full time series, right: zoom-in view, with corresponding spatial profile of the relative fraction $f_{rs}^l$ shown in (c) and dashed lines marking the snapshots times.
    (b) Time series for a large system ($L=32$). Left: whole system, right: a subsystem with edge length $9$.
    (d) Fourier components of $G_{rr}$ (rescaled by system size $N=L^2$) for small and large systems. Dashed line indicates oscillation frequency.
    (e) Two-time auto-correlation $G_{rr}$ and cross-correlation $G_{rs}$. Parameters for (a)-(g): $\Delta_r=3$.}
	\label{fig:vdW}
\end{figure}

Similar to LC, quasicycles exhibit a non-zero phase shift between the two components [Fig.~\ref{fig:all_to_all}(d)], indicating that quantum fluctuations preserve the predator-prey relation of LCs. The scaled Fourier components $N\mathcal{F}_\omega^{r}$ are displayed in Fig.~\ref{fig:all_to_all}(f). Their data collapse for different system sizes confirms the amplitude scaling $\delta n_{s,r}\propto1/\sqrt{N}$, indicating that the quasicycles are driven by finite-size noise~\cite{mckane2005predator,bressloff2010metastable,mckane2007amplified,boland2008how}.
Meanwhile, we observe an intrinsic frequency [$\omega \approx 2.4$, indicated by the gray dotted line] that is distinct from the one induced by Gaussian noise [$\Im[\lambda_0] \approx 3.7$ indicated by the blue lines in Fig.~\ref{fig:MF}(b)]. This difference arises from quantum fluctuations, a feature that underscores a different mechanism of noise amplification.

The all-to-all coupling stabilizes global oscillations against quantum jumps. This contrasts sharply with systems governed by vdW interactions, where local fluctuations cause desynchronization.
Consequently, global oscillations in the average Rydberg populations $n_s$ and $n_r$ are more pronounced in smaller systems ($L=8$) than in larger ones ($L=32$), as shown in Figs.~\ref{fig:vdW}(a) and (b).
In small systems  ($L=8$), short-time cyclic dynamics [right panel of Fig.~\ref{fig:vdW}(a)] 
and an alternating spatial profile in relative fraction $f_{rs}^l$  [Fig.~\ref{fig:vdW}(c)] occur.
In contrast to the deterministic case [Figs.~\ref{fig:vdW_wo_jump}(a) and (b)], the oscillation amplitude scales as $1/\sqrt{N}$ [Fig.~\ref{fig:vdW}(d)]. Nevertheless, a subspace analysis (probing a $9 \times 9$ region within a $32 \times 32$ array) reveals that synchronization persists locally, even when global coherence is lost. This localized phase locking is clearly manifested in the right panel of Fig.~\ref{fig:vdW}(b). 

Although global oscillations are significantly suppressed in larger systems, where individual trajectories resemble a stochastic steady state [see left panel of Fig.~\ref{fig:vdW}(b)]. The hidden predator–prey relationship is revealed in the two-time cross-correlation functions $G_{rs}(t)$. 
As evidenced in Fig.~\ref{fig:vdW}(e), these correlation functions exhibit a phase shift and the intrinsic frequency characteristic of predator-prey dynamic in their short-time dynamics.
However, the long-time periodicity is significantly disrupted by quantum-jump induced desynchronization, merging the several sharp Fourier peaks [Fig.~\ref{fig:all_to_all}(e)] into a single broadened feature [Fig.~\ref{fig:vdW}(d)].

\section{Discussion and conclusion}
In this work, we demonstrate the rise of emergent predator-prey cycles in a driven-dissipative two-component atomic array. Although the predator-prey dynamics is a classical example of self-organization far from equilibrium, their microscopic mechanisms remain difficult to isolate and control in natural ecosystems. 
By establishing the explicit connections between microscopic parameters, such as laser detuning, Rabi frequencies, and atomic interaction strength, and the emergent properties (frequency and phase shift) of the predator-prey cycles, our work positions the two-component Rydberg atomic lattices as a highly controllable experimental platform for exploring self-organizing phenomena in nonequilibrium systems. While we realize predator-prey cycles through laser detuning asymmetry, unequal interaction strengths can similarly induce dynamical instability, and oscillations also emerge for other asymmetric parameter choices. 

In classical population dynamics, finite-system effects are typically introduced via multiplicative white Gaussian noises, which induce various noise-induced phenomena absent in MF models~\cite{biancalni2014noise,biancalani2017giant,david2018stochastic}. Prior to the onset of the LC phase, our system exhibits quasicycles driven by the amplification of intrinsic quantum noises from quantum jumps.
As finite-size effects, these quantum quasicycles are expected to disappear in the thermodynamic limit.
Crucially, unlike their classical counterparts~\cite{mckane2005predator}, they exhibit a distinct intrinsic frequency not captured by Gaussian-noise approximations.

By comparing systems with all-to-all and short-range vdW coupling, both with and without quantum noise, we identify long-range interactions as a crucial mechanism for stabilizing global oscillations against local quantum jumps. 
The inverse square-root scaling of the oscillation amplitude with system sizes and the presence of oscillatory subsystems show that large systems comprise independent oscillating parts~\cite{dobramysl2018stochastic}. This contrasts sharply with global oscillatory states driven by self-organized bistability, where dissipative first-order phase transitions induce transient long-range correlations through system-spanning avalanches~\cite{xiang2024self}. 

The absence of global synchronization in systems with finite-range interactions further distinguishes our predator-prey cycles from dissipative time crystals~\cite{kessler2021obs,xiang2024self,wu2024dissipative}. Although both arise from spontaneous time-translation symmetry breaking, dissipative time crystals typically exhibit global phase coherence. In our system, such global coherence is present only in the all-to-all interaction limit; finite-range interactions destroy it through quantum-jump induced desynchronization.

Because of this desynchronization, global oscillations in finite-range systems are observed only for moderate system sizes. For large systems, global quantities such as average populations become insufficient to capture the collective dynamics; instead, spatially resolved observables and correlation functions reveal the presence of local oscillations even when the global signal is suppressed. Nevertheless, the MF instability remains physically relevant: local oscillations persist in the stochastic finite-range system but are hidden at the global level.

The OSDTWA method employed in this work is approximate in the presence of nonlinear interactions, and its quantitative accuracy is less controlled in strongly correlated regimes where higher-order correlations beyond the second moment become significant. However, the continuous action of quantum jumps and dissipation provides a natural cutoff for such quantum correlations, suggesting that the method is applicable to the driven-dissipative regime considered here.

\begin{acknowledgments}
 We are deeply grateful to U. C. T\"auber, Thomas Pohl, C. S. Adams, and Yuechun Jiao for their insightful suggestions. The authors also acknowledge the computational resources provided by the High Performance Computing Center of Nanjing University.
\end{acknowledgments}

\section*{Funding}
This work is supported by the National Natural Science Foundation of China (12347102 and 12547183), the Natural Science Foundation of Jiangsu Province (BK20233001), the Quantum Science and Technology-National Science and Technology Major Project (2024ZD0300101) and the Fundamental and Interdisciplinary Disciplines Breakthrough Plan of the Ministry of Education of China (JYB2025XDXM502). 

\section*{Author contributions}
Z. B. and Y.-X. X. conceptualized and designed the research. 
Y.-X. X. performed the simulations, analyzed the data, and wrote the original draft. 
W. L. and Z. B. reviewed and approved the manuscript.  
Y.-Q. M. supervised the research. 
All authors participated in discussions and contributed to the editing of the manuscript.

%

\end{document}